\documentclass[10pt,aip,twocolumn,superscriptaddress,showpacs]{revtex4-1}
\usepackage{graphicx}
\usepackage{color,soul}
\usepackage{latexsym}
\usepackage{subfigure}
\usepackage{amsmath}

\begin{document}

\title{Strain engineered domain structure and their relaxation in perpendicularly magnetized Co/Pt deposited on flexible polyimide} %

\author{Esita Pandey}%
\affiliation{Laboratory for Nanomagnetism and Magnetic Materials (LNMM), School of Physical Sciences, National Institute of Science Education and Research (NISER), HBNI, P.O.- Jatni, 752050, India}%
\author{Braj Bhusan Singh}%
\affiliation{Laboratory for Nanomagnetism and Magnetic Materials (LNMM), School of Physical Sciences, National Institute of Science Education and Research (NISER), HBNI, P.O.- Jatni, 752050, India}%
\author{Purbasha Sharangi}%
\affiliation{Laboratory for Nanomagnetism and Magnetic Materials (LNMM), School of Physical Sciences, National Institute of Science Education and Research (NISER), HBNI, P.O.- Jatni, 752050, India}%
\author{Subhankar Bedanta}%
\email{sbedanta@niser.ac.in}
\affiliation{Laboratory for Nanomagnetism and Magnetic Materials (LNMM), School of Physical Sciences, National Institute of Science Education and Research (NISER), HBNI, P.O.- Jatni, 752050, India}%
\date{24 January 2020}%


\begin{abstract}
	The demand of fast and power efficient spintronics devices with flexibility requires additional energy for magnetization manipulation. Stress/and strain have shown their potentials for tuning magnetic properties to the desired level. Here, we report a systematic study for the effect of both tensile and compressive stresses on the magnetic anisotropy (MA). Further the effect of stress on the domain structure and magnetization relaxation mechanism in a perpendicularly magnetized Co/Pt film has been studied. It is observed that a minimal in-plane tensile strain has increased the coercivity of the film by 38$\%$ of its initial value, while a very small change of coercivity has been found under compressive strain. The size of ferromagnetic domains decreases under tensile strain, while no change is observed under the compressive strain. Magnetization relxation measured at sub-coercive fields yields longer relaxation time in the strained state.

\end{abstract}

\maketitle
 Miniaturization and flexibility with high speed and low power consumption are the key aspects for developing next generation spintronic devices \cite{tudu2017recent,bao2016flexible,sheng2018flexible}. Perpendicular magnetic anisotropic (PMA) systems such as Co/Pt bilayes have shown their importance in increasing thermal stability and developing non-volatile magnetic random access memories (MRAM's) with low current density requirement for magnetization switching at 20 nm bit size level \cite{tudu2017recent,ikeda2010perpendicular,gopman2016strain,sampaio1996magnetic,ikeda2007magnetic,lee2014thermally,vemulkar2016toward}. The major challenge in further downscaling of devices is the higher power dissipation \cite{barangi2015straintronics}. Various approaches have been taken to reduce the power consumption during magnetization switching e.g. spin transfer torque (STT), current induced domain wall (DW) motion etc. \cite{liu2012current,cui2018magnetization}. However, involvement of charge current is still a major factor to further reduce the power consumption. In this context, stress controlled method has been proven to be a promising one \cite{biswas2017experimental,bukharaev2018straintronics}. Strain engineering became an appealing approach to meet the current device requirement. There are various possible ways to generate strain on a thin film deposited on rigid substrate e.g. by lattice mismatch \cite{chen2004enhanced,abadias2018stress}, voltage application via transducers \cite{lei2011magnetization} etc., whereas for a film deposited on flexible substrate bending, stretching, peeling or twisting mechanism also can generate adequate stress \cite{harris2016flexible}. Such mechanical methods transfer almost uniform stress from substrate to the film, which gives rise to device flexibility. In recent years the stress effect on magnetic anisotropy and domain structure in various magnetic films with in-plane MA have been studied \cite{chen2009effects,sheng2018flexible}. However, the stress effect in flexible magnetic films with  PMA is very less explored. Theoretical studies predicted that magnteic anisotropy energy as well as the easy axis of Fe/Pt multilayers can be modified by strain  \cite{tao2017engineering}. Shepley $et$ $al$. showed that the MA of a Co/Pt film decreased with increasing compressive strain while the DW velocity was increased by 30 to 100$\%$ \cite{shepley2015modification}. Later the stress tuned DW velocity of Co/Pt depends on DW formation energy (balance of anisotropy and exchange energy) is also reported \cite{shepley2018domain}.  However, a systematic study on the effect of different types of strain e.g. tension, compression, peeling etc. on the MA, domain nucleation, its structure along with relaxation mechanism is still lacking. Exploring the effect of such magneto-mechanical coupling on different magnetic properties will not only help to understand the basics but also holds importance from application viewpoint. Here, we report a comprehensive study of the effect of tensile and compressive strain on the MA, domain nucleation and relaxation mechanism in a perpendicularly magnetized Co/Pt film prepared on polyimide substrate.

\begin{figure}[h!]
	\centering
	\includegraphics[width=0.9\linewidth]{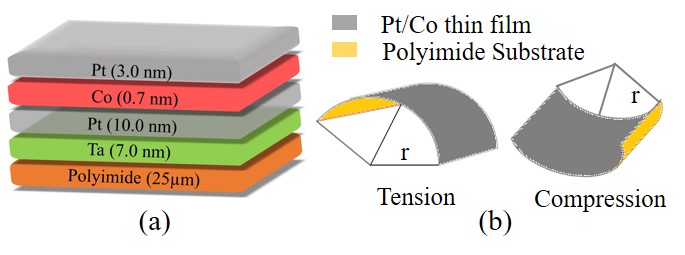}
	\caption{Schematic of (a) Co/Pt thin film deposited on flexible polyimide substrate, (b) Tensile and compressive strain generation by bending of the film achieved by convex and concave shaped molds, respectively.}
	\label{fig1}
\end{figure}

We have prepared Pt/Co thin film upon 25 $\mu$m thick polyimide (PI) substrate by dc magnetron sputtering in a high vacuum chamber manufactured by Mantis Deposition Ltd, UK. Polyimide (PI) substrate was cleaned ultrasonically by isopropanol before deposition. The sample structure is the following:
PI/Ta(7nm)/Pt(10nm)/Co(0.7nm)/Pt(3nm) as shown in fig.\ref{fig1}(a). The base pressure of the deposition system was better than 1$\times$10$^{-7}$ mbar. The rate of deposition for Co, Pt and Ta layers were 0.10 Å/s, 0.15 Å/s and 0.13 Å/s respectively. The thickness was measured by a quartz crystal monitor (QCM) mounted close to the substrate holder. Substrate was rotated at 10 rotation per minute (rpm) during deposition to get uniformity in the samples. A 7 nm thick Ta buffer layer was used for better adhesion and growth of Pt layer along (111) direction. A 3 nm thick Pt was used as a capping layer to prevent the oxidation of Co layer. 

Magnetic properties of the sample have been recorded at both the flat as well as bend state. Different tensile (compressive) strain has been generated on the film by sticking it on convex (concave) mold made of aluminium with different radius of curvature. Schematic of both mold is shown in fig.\ref{fig1}(b) and the molds used in the measurement are shown in supplementary fig.S1. Domain nucleation and propagation at both states of the sample have been recorded by polar magneto optic Kerr effect (PMOKE) based microscope manufactured by Evico Magnetics, Germany. Further, relaxation measurements have been carried out in the flat as well as bend state of the film.

\begin{figure}[h!]
	\centering
	\includegraphics[width=.9\linewidth]{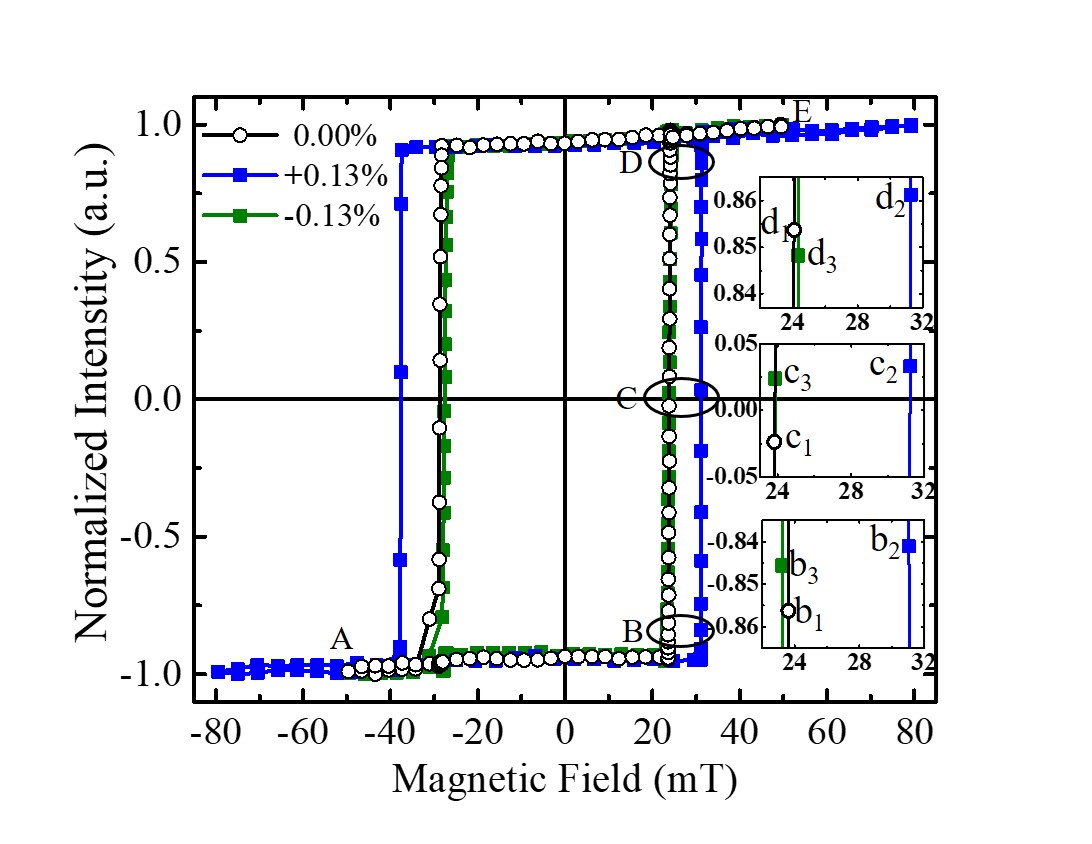}
	\caption{Hysteresis loops measured by PMOKE microscopy at unstrained (circle) and strained (square) states of the sample. Inset: Zoom in view of the black marked regions (B-D) of hysteresis loops, at which domain images are shown in fig.\ref{fig4}.}
	\label{fig2}
\end{figure}

We have measured magnetic hysteresis loops in the polar mode by bending the sample towards inward and outward direction. Such bending induced strain on the film can be determined as,
\begin{equation}
\epsilon=\frac{t}{2r \pm t}
\label{q1}
\end{equation} 
where $\epsilon$ is the applied strain, `+' sign for the tensile strain, `-' sign is for the compressive one,  $r$ is the radius of the mold used and $t$ is the total thickness of the film and substrate \cite{qiao2017enhanced,dai2012mechanically,wang2005magnetostriction}.Using molds almost uniform strain is possible to transfer from the substrate to the thin film (as thickness of the film is very low in comparison to the substrate). Coercivity ($H_C$) of the sample was found to be $26.9$ mT in flat state which increases (decreases) to $34.4$ ($25.7$) mT under 0.13 $\%$ tensile (compressive) strain (fig.\ref{fig2}). Even a minimal in-plane tensile strain ($<0.1\%$) has increased the coercivity largely ($\sim10$ mT) whereas the effect of compression is almost negligible. Such behaviour can be explained by comparing the strength of stress induced anisotropy over interfacial anisotropy. Magneto-elastic energy of a thin film can be expressed as:
\begin{equation}
E_{ME}=\frac{3}{2}\lambda \sigma\sin^2\theta  
\label{q2}
\end{equation}
where $\lambda$  is co-efficient of magnetostriction, $\sigma$ is the applied stress, $\theta$ is the angle between the stress and magnetization axis \cite{cullity2009introduction}. As reported earlier  $\lambda$ is negative for Co/Pt system along (111) direction \cite{shepley2015modification} while the applied stress $\sigma$ is positive (negetive) for tensile (compressive) strain, hence the $\lambda$$\sigma$ product is negative (positive) which induces an easy axis along the perpendicular plane (in-plane) of the sample to minimize total energy of the system. As the interfacial and tensile (compressive) stress induced MA acted along the same (opposite) direction, hence it is easier (harder) for tension (compression) to increase (decrease) the overall MA. Hence the strength of magneto-mechanical coupling effect require to modify MA significantly is different, depending upon the type of strain. 

\begin{figure}[h!]
	\centering
	\includegraphics[width=.8\linewidth]{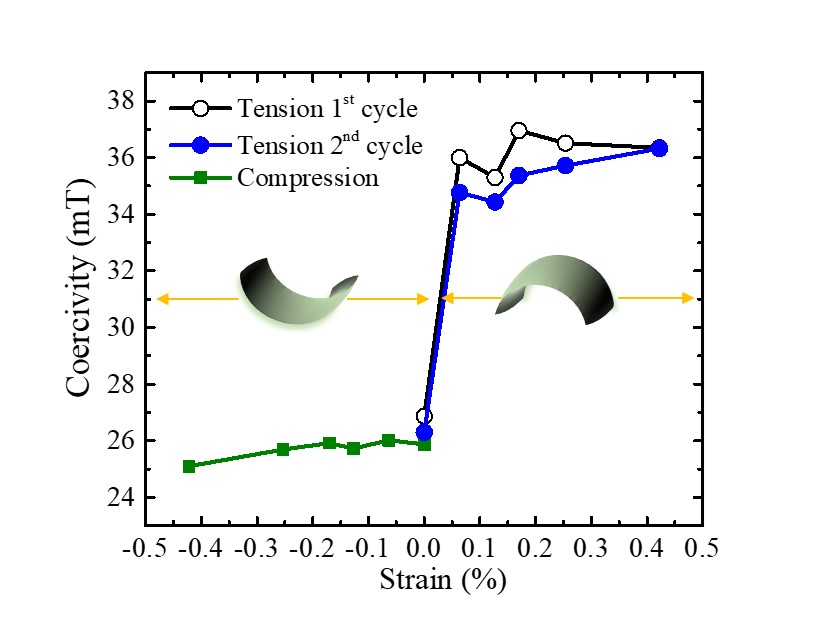}
	\caption{Coercivity change for 1$^s$$^t$ (black circle) and 2$^n$$^d$ (blue circle) bending cycles under tensile strain. 2$^n$$^d$ bending cycle is measured after keeping the sample under $\sim$0.5$\%$ tensile strain for 15 days. The square  symbols represents the chnage in coercivity by applying compressive strain.   }
	\label{fig3}
\end{figure}

The variation of coercivity with increasing tensile/compressive strain has been plotted in fig.\ref{fig3}. Here the 1$^s$$^t$ cycle represents the values of $H_C$  obtained from the hysteresis when the sample is bent for the first time at different radii after deposition. After that we kept the sample under 0.5$\%$ tension for 15 days and re-measured $H_C$ at different strain values which is plotted as 2$^n$$^d$ cycle. The nature of the two graphs indicates an excellent endurance of the film under long term stress application. However under compression the effect is almost negligible and $H_C$ decrease by $\sim$1.0 mT when a large ($\sim$0.5$\%$) strain is applied. Origin of such different behaviour is already explained in the previous section.

\begin{figure}[h!]
	\centering
	\includegraphics[width=1\linewidth]{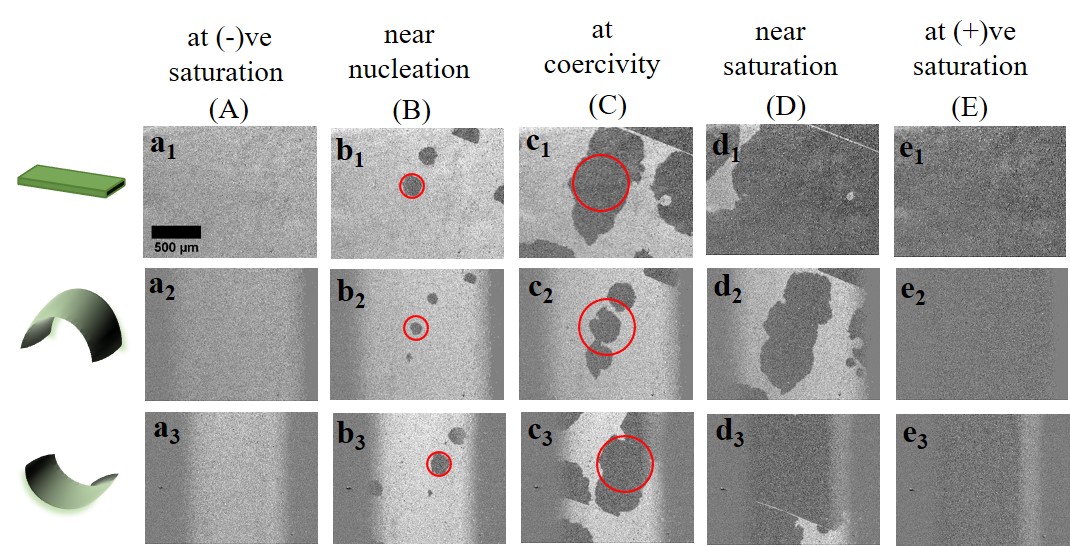}
	\caption{Domain images recorded at negative saturation (A), near nucleation (B), at coercivity (C), near positive saturation (D) and at positive saturation (E) points of the hysteresis loops (fig.\ref{fig2}) by MOKE microscope at flat and bend ($\pm$ 0.13$\%$ strain) state of the sample. Scale bar is $500$ $\mu$m for all the domain images.}
	\label{fig4}
\end{figure}

Bubble domain has been observed at the initial flat state of the sample, as the anisotropy ratio, $Q=K_{u}/K_{d}\gg$1, where $K_{u}$ is the uniaxial anisotropy and $K_{d}$ is the stray field energy density, respectively \cite{hubert1998magnetic}. The domain images are recorded at saturation, near nucleation and sub-coercive field values of each hysteresis loops as shown in fig.\ref{fig4}. All the domain images are captured at the same place of the sample, to easily compare them under different strained conditions. Red marked circles in fig.\ref{fig4} (b1 and c1) are showing the domain size for better visualization in the flat state of the sample. Same size of circles are drawn on the domain images recorded under tensile (b2 and c2) and compressive (b3 and c3) strained states to quantify the  contraction/elongation of the domains in comparison to the flat state.  It is observed that application of an in-plane tensile strain reduces the size of bubble domains significantly. As $H_C$ of the sample has increased by 38$\%$ under tension, it hindered the DW propagation which reduces the domain size.  Magnetization reversal alongwith domain nucleation also takes place at higher (lower) field values due to the increases (decreases) in overall energy barrier under tensile (compressive) strain. The difference between nucleation field ($H_N$) and $H_C$  was found to be 2.5 mT at the flat state whereas it increases (decreases) to 4 ($\sim$1.2) mT under tension (compression). This can be a result of increase(decreases) in Bloch wall formation energy due to the appliaction of +0.13$\%$ (-0.13$\%$) strain. The effect of -0.13$\%$ strain is not much visible possibly due to its less effect on total anisotropy field. Under in-plane tensile strain the DW motion is expected to be reduced due to increase in potential barrier for the magnetization reversal. Shepley $et$ $al$. has already reported a faster DW motion of Pt/Co/Pt tri-layer system under out of plane tensile strain where the sample was prepared near to spin-reorientation transition (SRT) region. As our sample is far from SRT so the effect of compression is very less in our case.

As the compressive strain does not have much effect in modifying magnetic properties of our film hence we performed the relaxation measurement under tensile stress only and compared with the flat one. Under the application of a constant Zeeman energy to the system, the magnetization relaxation can be recorded with respect to time. This measurement basically reveals the efficiency of thermal activation energy to complete magnetization reversal via domain nucleation and DW motion. We have performed similar measurement on our film by PMOKE based microscopy. The sample was first saturated at a negative saturation field, then the field was increased manually near to positive sub-coercive field values e.g. 0.95 $H_{C}$, 0.97 $H_{C}$ and kept it constant there.
 In this scenario, further reversal takes place with the help of thermal energy. Domain images were captured during the whole thermal relaxation process. The amount of dark grey contrast represents a measure of the magnetization for the sample. The intensity of each image was extracted by ImageJ software and the average intensity of all images captured at each second were calculated. The normalized intensity is then plotted against the time taken to complete the reversal. Here the normalized intensity as a function of time basically reflects the net magnetization relaxation of the sample. Amongst various proposed models to explain magnetization relaxation phenomena Fatuzzo-Labrune model is extensively used for FM thin films \cite{labrune1989time,xi2008slow,adjanoh2011compressed,chowdhury2016study,mallick2015effect,mallick2015size}. However, the approximation of single energy barrier made in this model does not holds for a real thin film which consists of a distribution of energy barrier with defects and inhomogeneity throughout the surface. Hence to fit our experimental data we have used the compressed exponential function \cite{xi2008slow,mallick2018relaxation},

  \begin{equation}
  I(t)=I_{1} +I_{2}(1-exp(-(\frac{t}{\tau})^\beta))
  \label{q3}  
  \end{equation}
  
  \begin{figure*}
  	\centering
  	\includegraphics[width=0.7\linewidth]{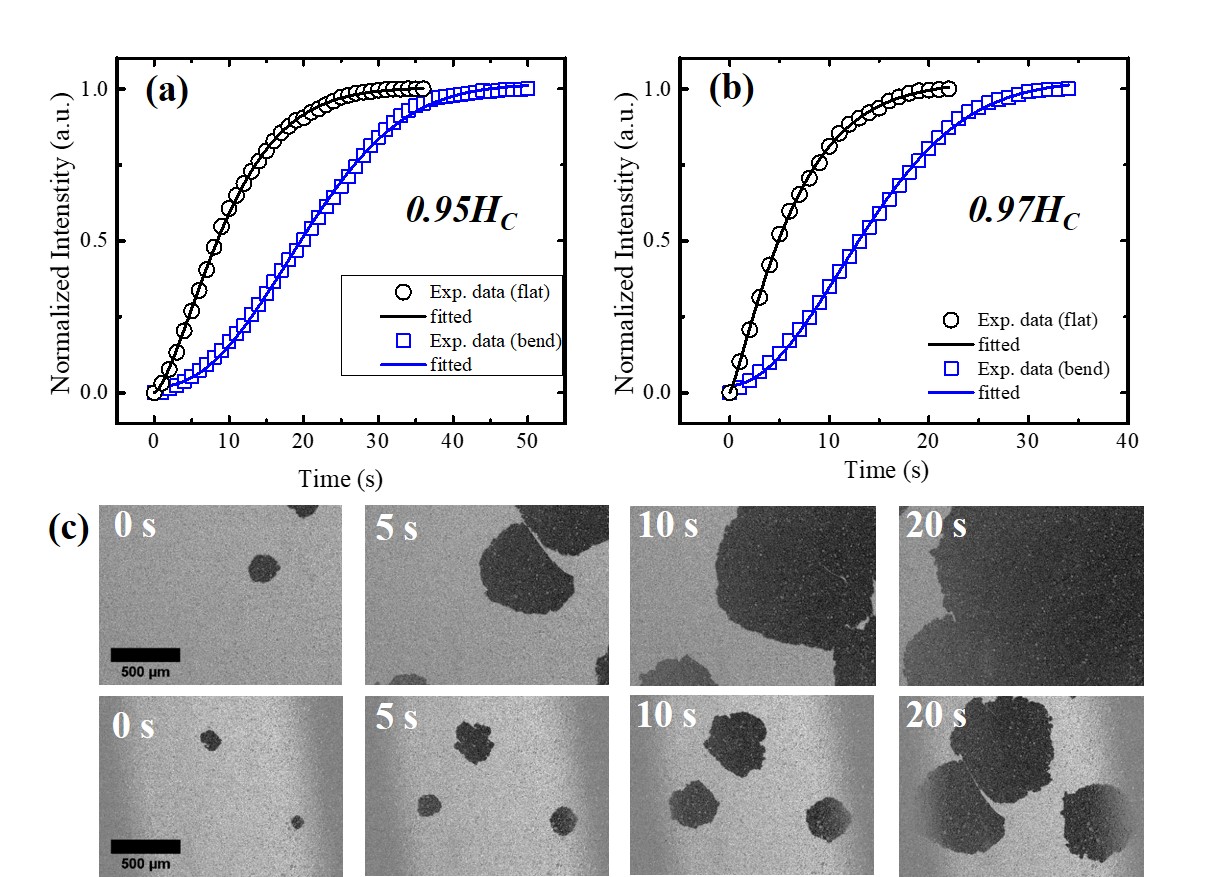}
  	\caption{Relaxation measurement performed on Co/Pt sample at two sub-coercive field values (a) 0.95 $ H_{C}$ and (b) 0.97 $ H_{C}$, at 0.13$\%$ strained (blue curve) and unstrained (black curve) state. The measured data were well fitted by the compressed exponential function.(c) Domain expansion by thermal energy at constant (0.95 $H_{C}$) magnetic field recorded by PMOKE microscope at both unstrained (upper panel) and +0.13$\%$ tensile strained state (lower panel), during different time interval (0, 5, 10 $\&$ 20 s) of the reversal process. Scale bar is 500 $\mu$m for all domain images.}
  	\label{fig5}
  \end{figure*}
  where $I(t)$ is the Kerr intensity measured at time t, $I_{1} + I_{2}$ is the normalized Kerr intensity, $\beta$ is an exponent varying within 1 (nucleation dominated) to 3 (DW motion dominated) and $\tau$ is the relaxation time constant.
  At the flat state the relaxation happens via both domain nucleation as well as DW motion ($\beta$ = 1.46$\pm$0.01 and 1.19$\pm$0.01) and the relaxation time constant is found to be 10.90 $\pm$0.04 and 6.86$\pm$0.07 s at 0.95 and 0.97 $H_C$, respectively. However, a slower relaxation phenomenon occurs in presence of $ 0.13\%$ tensile strain which started via domain nucleation and dominated further by DW motion ($\beta$= 2.23$\pm$0.04 and 1.95$\pm$0.04). Relaxation time constant at the bend state was 23.63$\pm$0.14 and 16.24$\pm$0.13 s at 0.95 and 0.97 $H_{C}$, respectively. Nature of the relaxation curve also reflects a slower relaxation phenomenon as shown in fig.\ref{fig5} (a-b). This is supported by a larger relaxation time constant ($\tau$) extracted from the fitting. To explain the origin behind this slow relaxation we have analyzed the domain images correspond to 0.95 $H_{C}$ of both states at different time intervals viz. 0, 5, 10 and 20 s (fig.\ref{fig5}.c). Here 0 s signifies the initial time when the first image was taken after giving a constant (0.95 $ H_{C}$) magnetic field. As 0.95 $  H_{C}$ field is greater than the nucleation field for both the states, hence bubble domains are visible at both the images of 0 s. Further all the images were taken at the same place of the sample hence at 0 s the size of the domain in the bent state is smaller due to obvious reason. Further with increasing time (5, 10 and 20 s) it has been found that the domain propagates slower in the bend state in comparison to the flat one due to its reduced wall motion. This reduced wall motion also reflects in the difference between $H_{N}$  and $H_{C}$ field values of both state, as discussed earlier. In about 20 s of the relaxation process, flat  sample almost completes the reversal while for the bend one, it is still far away from saturation. From the above result we expect that a Co/Pt thin film prepared near to SRT will show a significant reduction of relaxation time under compressive strain, which holds its potential in switching application.
  
  We presented the effect of tensile/compressive strain on the magnetization reversal, magnetic domains and relaxation mechanism of a Co/Pt flexible film. It is observed that tensile strain has largely modified the MA, domain nucleation and relaxation mechanism of our film, whereas the effect of compression is found to be less prominent.
  A substantial increase in MA, nucleation field and decrease in domain size attributed mainly to the magneto mechanical coupling effect. In spite of this high coupling effect, our flexible film exhibits excellent endurance for long term stress application. A longer field span required to complete the magnetization reversal at the strained state, reflects a increase in potential barrier for the DW movement which modifies the relaxation time of the film. Relaxation measurements indicates $\sim$13 sec of relaxation time  difference  between  the unstrained and curved state which is further endorsed by comparing domain images taken at different interval of time. This study further demands to explore the effect of compression on the relaxation time (switching time) of a Co/Pt film prepared near to SRT region. It is expected that in such scenario the relaxation will be faster which has significant potential for faster device application.
  
  Authors would like to thank Mr. Brindaban Ojha and Dr. Tanmoy Chabri for their help during the relaxation measurements. We also thank the Department of Atomic Energy (DAE) of the Govt. of India for providing the funding to carry out the research. BBS acknowledges DST for INSPIRE faculty fellowship.

 \section*{Supplementary Material}
 {See  supplementary  material  for Figure S1 which shows the image of the convex and concave shaped molds used in the measurement.
 }\\
 
\bibliography{references}
\end{document}